\documentclass[12pt]{article}
\usepackage{amsmath,epsfig}

\begin{document}

\begin{center}
{\Large \bf Analyzing spatial coherence using a single mobile field sensor}\\
\medskip
P. A. Fridman\\
ASTRON,
Oude Hoogeveensedijk 4, 7991PD Dwingeloo, The Netherlands
\vspace{5mm}
\end{center}

\begin{abstract}
 According to the Van Citter-Zernike theorem the intensity distribution of a spatially incoherent source and the mutual coherence function of the light impinging on  two wave sensors are related.  It is the  comparable relationship using a single mobile  sensor moving at a certain velocity relative to the source which  is calculated in this article. The autocorelation function of the electric field  at the sensor contains information about  the intensity distribution. This expression could  be employed in aperture synthesis.

\end{abstract}
\section{Introduction}
The mutual intensity and degree of coherence for light from an extended incoherent quasi-monochromatic source is explained by the Van Citter-Zernike theorem \cite{ref1}. This theorem is widely applied  in optics and radio domains and in  aperture synthesis, in particular\cite{ref2}.
Fig 1a shows the setup used in \cite{ref1}  to illustrate the Van Citter-Zernike theorem. A screen $\cal{A}$ is illuminated  by an extended quasi-monocromatic spatially incoherent source $ \sigma .$  The source $ \sigma $  occupies a fragment of a plane parallel to  screen $\cal{A}$.  The linear dimensions of the source are small compared to the distance $R$ between the source and the screen.   The medium between the source and the screen is homogeneous and  $\ v$\ is \ the velocity of light in the medium.  Two points $P_{1}$ and $P_{2}$ \ are chosen on  screen $\cal{A}$  and the mutual intensity of  light emitted by a source point $S$  is calculated for these points. The angles between $OO^{'}$ \ and the lines connecting  point $S$ and \ $P_{1}$ and $P_{2}$ are small.  The wave field created by  $S$ is represented by a complex analytical signal  $s(t)=E(t)\exp \{i[\Phi (t)-2\pi \overline{\nu }t]\},$ the envelope $E(t)$ and the phase $\Phi (t)$  vary slowly in comparison with $\cos (2\pi \overline{\nu }t)$ and $\sin (2\pi \overline{\nu }t),$ $\overline{\nu }$ is the mean frequency. The complex envelope $A(t)=E(t)\exp [i\Phi (t)]$ will be employed later. The frequency interval  of width  $\Delta \nu $ is small compared to the mean frequency  $\overline{\nu} .$ Under all these assumptions the  mutual intensity, or the spatial coherence function of the field $s(t)$ is \cite{ref1} :
\begin{equation}
J(P_{1},P_{2})=\int\limits_{\sigma }I(S)\frac{\exp \{i\overline{k}[R_{1}(S)-R_{2}(S)]\}}{R_{1}(S)R_{2}(S)}dS,
\end{equation}
where $R_{1}(S)$ and $R_{2}(S)$ denote the distance between a typical source point S and the points $P_{1}$ and $P_{2}$ and \ $\overline{k}=2\pi \overline{\nu }/v=2\pi /\overline{\lambda }$ is the wave number of the medium,  $\overline{\lambda }$ is the mean wavelength and  $I(S)$ is the intensity per unit area of the source. Eq. (1)  is invertible, within reasonable limits, and is the basic  equation in  aperture synthesis \cite {ref2}. In the simplest case the first  wave sensor (receiver) is positioned at  the fixed point $P_{1}$, while the second sensor is positioned in series \ at $P_{2}$ \ and \ other points so as to obtain several samples of the spatial coherence function. Details of the proof of  Eq. (1) are not included at this point but they will be reproduced in the calculation of  mutual intensity for the setup in Fig. 1b which is the objective of this article.

\begin{figure}[t]
\centerline{\includegraphics[width=10.0cm,height=10.0cm]{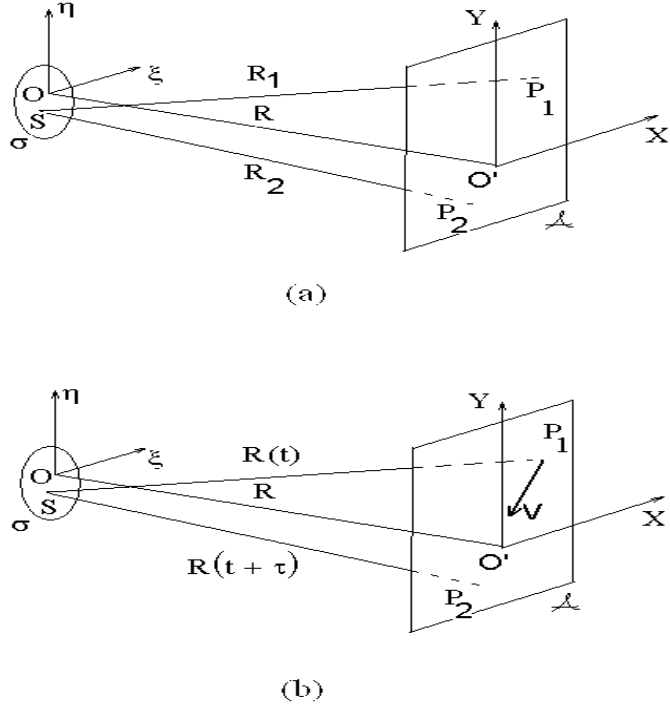}}
\caption{Setup  for the  Van Citter-Zernike theorem for: a) two sensors; b) one moving sensor.}
\end{figure}

\section{Spatial coherence measured on a moving receiver}

The difference between Fig. 1a and b is the following. A receiver in Fig. 1b is initially placed at point $P_{1}$ and moves in the direction of point $P_{2}$ with  velocity $V$. The signal $s(t)$ is processed in {\it real time}, or recorded  and later processed {\it off-line }with the aim of  calculating the autocorrelation function
\begin{equation}
J(P,\tau )=\left\langle s(t)s^{\star }(t+\tau )\right\rangle ,
\end{equation}
where the sharp brackets denote the time average and  the raised asterisk indicates the complex conjugate.  Now $J(P,\tau )$ can be calculated using the  approach  applied in \cite{ref1}  for Eq. (1).
The source is divided into  $M$ elements $d\sigma _{1},d\sigma _{2}...d\sigma _{M}$ \ centered on points $S_{1},S_{2},...S_{M},$ the linear dimensions of the elements are smaller than the mean wavelength $\overline{\lambda }.$ Let $s_{m}(t)$ and $s_{m}(t+\tau )$ be the complex signals at point $P_{1}$ at moment $t$ \ and \ at the receiver at moment $t+\tau $ when it reaches \ point $\ P_{2}$ moving with velocity $V,$ respectively. The total \ signals at these moments \ are
\begin{equation}
s(t)=\sum\limits_{m}s_{m}(t),s(t)=\sum\limits_{m}s_{m}(t+\tau),
\end{equation}

and the correlation function is
\begin{equation}
J(P,\tau )=\left\langle s(t)s^{\star }(t+\tau )\right\rangle =\sum\limits_{m}\left\langle s_{m}(t)s_{m}^{\star }(t+\tau )\right\rangle +\sum \sum_{m\neq n}\left\langle s_{m}(t)s_{n}^{\star }(t+\tau )\right\rangle
\end{equation}
The second sum is equal to zero due to the incoherence of the signals from different source elements $s_{m}$ and \ $s_{n},m\neq n.$ \ Let $R_{m,P_{1}}$ and \ $R_{m,P_{2}}(\tau )$ be the distances of the source element $d\sigma _{m}$ from  points $P_{1}$ and $P_{2},$ respectively.  So the partial signals are

\begin{eqnarray}
s_{m}(t)=A_{m}[t-\frac{R_{m,P_{1}}}{v}]\frac{\exp [-i2\pi \overline{\nu }(t-\frac{R_{m,P_{1}}}{v})]}{R_{m,P_{1}}},\nonumber\\
s_{m}(t+\tau )=A_{m}[t+\tau -\frac{R_{m,P_{2}}(\tau )}{v}]\frac{\exp \{-i2\pi \overline{\nu }[t+\tau -\frac{R_{m,P_{2}}(\tau )}{v}]\}}{R_{m,P_{2}}(\tau )}
\end{eqnarray}
The averaged product $\ \left\langle s_{m}(t)s_{m}^{\star }(t+\tau )\right\rangle $ \ is
\begin{equation}
\left\langle s_{m}(t)s_{m}^{\star }(t+\tau )\right\rangle =\left\langle A_{m}(t)A_{m}^{\star }(t+\tau -\frac{\Delta R_{m}(\tau )}{v})\right\rangle \frac{\exp \{^{i2\pi \overline{\nu }[\tau -\frac{\Delta R_{m}(\tau )}{v}]}\}}{R_{m,P_{1}}R_{m,P_{2}}(\tau )}
\end{equation}
where $\Delta R_{m}(\tau )=R_{m,P_{1}}-R_{m,P_{2}}(\tau ).$ The first factor in Eq. (6) is a complex correlation function of the envelope $A_{m}(t),$ which will be denoted as $C_{m}[\tau -\frac{\Delta R_{m}(\tau )}{v}]$. Quantity $\left\langle s_{m}(t)s_{m}^{\star }(t+\tau )\right\rangle $ is the correlation function corresponding to the source element $d\sigma _{m}.$ Integration over the whole source $\sigma $ , i.e., \ the transition to  continuum in Eq. (4) gives
\begin{equation}
J(P,\tau )=\int\limits_{\sigma }I(S)C[\tau -\frac{\Delta R(\tau ,S)}{v}]\frac{\exp \{i2\pi \overline{\nu }[\tau -\frac{\Delta R(\tau ,S)}{v}]\}}{R_{P_{1}}(S)R_{P_{2}}(S,\tau )}dS.
\end{equation}
Let $(\xi ,\eta )$ be the coordinates of a source point $S$, refered to axes at $O$, and let $(X_{0},Y_{0})$ and $(X_{\tau },Y_{\tau })$ be the coordinates of $P_{1}$ and $P_{2}(\tau )$, refered to the parallel axes at $O^{^{\prime }}$, Fig.1b. The distance $OO^{^{\prime }}$ is equal to $R$. Then
\begin{equation}
R_{P_{1}}(S)=\sqrt{(X_{0}-\xi )^{2}+(Y_{0}-\eta )^{2}+R^{2}}\simeq R+\frac{(X_{0}-\xi )^{2}+(Y_{0}-\eta )^{2}}{2R.}
\end{equation}
The same approximation is valid for $R_{P_{2}}(S,\tau )=R+\frac{(X_{\tau }-\xi )^{2}+(Y_{\tau }-\eta )^{2}}{2R}$ and the difference $\Delta R(\tau ,S)$ is
\begin{equation}
\Delta R(\tau ,\xi ,\eta )\simeq \frac{(X_{0}^{2}+Y_{0}^{2})-(X_{\tau }^{2}+Y_{\tau }^{2})}{2R}-\frac{(X_{0}-X_{\tau })\xi +(Y_{0}-Y_{\tau })\eta }{R}.
\end{equation}
$R_{P_{1}}(S)$ and $R_{P_{2}}(S,\tau )$ in the denominator of the integral in Eq. (7) may be approximated by $R.$ Using the notations
\begin{equation}
\frac{X_{0}-X_{\tau }}{R}=p_{\tau },\frac{Y_{0}-Y_{\tau }}{R}=q_{\tau },\psi (\tau )=2\pi \overline{\nu }\tau +\frac{\overline{k}[(X_{0}^{2}+Y_{0}^{2})-(X_{\tau }^{2}+Y_{\tau }^{2})]}{2R},
\end{equation}
we get
\begin{equation}
J(P,\tau )=\frac{e^{i\psi (\tau )}}{R^{2}}\int\limits_{\sigma }I(\xi ,\eta )C[\tau -\frac{\Delta R(\tau ,\xi ,\eta )}{v}]\exp [-i\overline{k}(p_{\tau }\xi +q_{\tau }\eta )]d\xi d\eta .
\end{equation}

Analyzing \ Eq.  (11)  we see that it represents a Fourier transform \ of the intensity function \ of the source (as in the  Van Citter-Zernike theorem),  but there is a weighting factor C($\tau )$ in the integral which must be taken into account when an inverse Fourier transform is performed  to find $I(\xi ,\eta ).$ Another substantial feature is that with a moving receiver we get a  continuum of baselines between points $P_{1}$ and $P_{2}$, i.e.,  there is a  scan of baselines in the interval  [0,0;$p_{\tau },q_{\tau }].$ In theory, it is possible  to capitalize on this property during  aperture synthesis procedure \cite{ref3},\cite{ref4}.






\end{document}